\documentclass[journal=jacsat,manuscript=article,layout=twocolumn]{achemso}
\usepackage{amsmath}
\usepackage{amsfonts}
\usepackage{graphicx}
\usepackage{physics}
\usepackage{siunitx}
\usepackage[nameinlink,capitalize]{cleveref}
\usepackage{color}
\usepackage[T1]{fontenc}

\renewcommand{\vec}[1]{\boldsymbol{\mathbf{#1}}}
\def\TM{\mathrm{TM}}
\def\TE{\mathrm{TE}}
\def\LcTE{L_b^{(\TE)}}
\def\LcTM{L_b^{(\TM)}}

\author{Félicien Appas}
\affiliation{Université Paris Cité, CNRS, Laboratoire Matériaux et Phénomènes Quantiques, 75013 Paris, France}
\alsoaffiliation{Current address: ICFO - Institut de Ciencies Fotoniques, The Barcelona Institute of Science and Technology, Castelldefels (Barcelona) 08860, Spain}
\altaffiliation{These authors contributed equally to this work}
\author{Othmane Meskine}
\affiliation{Université Paris Cité, CNRS, Laboratoire Matériaux et Phénomènes Quantiques, 75013 Paris, France}
\altaffiliation{These authors contributed equally to this work}

\author{Aristide Lemaître}
\affiliation{Université Paris-Saclay, CNRS, Centre de Nanosciences et de Nanotechnologies, 91120, Palaiseau, France}
\author{José Palomo}
\affiliation{Laboratoire de Physique de l’École normale supérieure, ENS, Université PSL, CNRS, Sorbonne Université, Université Paris Cité, F-75005 Paris, France}
\author{Florent Baboux}
\affiliation{Université Paris Cité, CNRS, Laboratoire Matériaux et Phénomènes Quantiques, 75013 Paris, France}
\author{Maria I. Amanti}
\affiliation{Université Paris Cité, CNRS, Laboratoire Matériaux et Phénomènes Quantiques, 75013 Paris, France}
\email{maria.amanti@u-paris.fr}
\author{Sara Ducci}
\affiliation{Université Paris Cité, CNRS, Laboratoire Matériaux et Phénomènes Quantiques, 75013 Paris, France}
\email{sara.ducci@u-paris.fr}

\title{Broadband biphoton generation and polarization splitting in a monolithic AlGaAs chip}

\begin{document}

\begin{abstract}
	The ability to combine various advanced functionalities on a single chip is a key issue for both classical and quantum photonic-based technologies. 
	On-chip generation and handling of orthogonally polarized photon pairs, one of the most used resources in quantum information protocols, is a central challenge for the development of scalable quantum photonics circuits; in particular, the management of spectrally broadband biphoton states, an asset attracting a growing attention for its capability to convey large-scale quantum information in a single spatial mode, is missing.
	Here, we demonstrate a monolithic AlGaAs chip including the generation of broadband orthogonally polarized photon pairs and their polarization splitting; \SI{85}{\percent} of the pairs are deterministically separated by the chip over a \SI{60}{\nano\metre} bandwidth. The quality of the two-photon interference at the chip output is assessed via a Hong-Ou-Mandel experiment displaying a raw visibility of \SI{75.5}{\percent} over the same bandwidth. These results, obtained for the first time at room temperature and telecom wavelength, in a platform combining strong confinement, high second-order nonlinearity, electro-optic effect and direct bandgap, confirm the validity of our approach and represent a significant step towards miniaturized and easy-to-handle photonic devices working in the broadband regime for quantum information processing.
\end{abstract}

\section{Introduction}

Photonic quantum technologies are a valuable approach for quantum information processing including secure communications, quantum computation, simulation and metrology~\cite{Flamini2018,Polino2020}. 
While quantum photonic experiments in bulk optics continue to produce important results demonstrating the advantage of quantum technologies over classical ones in several domains~\cite{HanSen2020,Madsen2022,Tse2019,Collaboration2019},
integration at the chip scale has become essential to scale up these concepts and transform laboratory demonstrators into real-world technologies \cite{Wang2020,Pelucchi2022}.
In the last two decades, significant advances on integrated quantum photonic circuits have enabled the generation and manipulation of quantum states of light at an increasing scale and level of complexity exploiting a variety of optical material platforms. Most of these achievements have been done using glass or indirect-bandgap $\chi^{(3)}$ materials such as silicon nitride, silicon-on-insulator or silica on silicon \cite{Yunhong2021}, despite the fact that the use of $\chi^{(2)}$ materials leads to more efficient frequency conversion processes and allow for easy pump filtering. Important results using this second class of materials have been obtained with lithium niobate-based photonics circuits, having led to the demonstration of an on-chip squeezing experiment~\cite{Mondain19,Nehra2022}, a two-channel spectrally degenerate polarization entangled source~\cite{Sansoni2017} and  a nonlinear integrated quantum electro-optic device~\cite{Luo2019}. However, in the first two works the chip doesn’t manage polarization, while in the third and fourth ones the devices work in a narrow spectral bandwidth regime and have typical sizes of several centimeters. Polarization handling of broadband biphoton states on miniature chips would constitute a crucial resource in the toolbox of integrated quantum photonic circuits. Indeed, on the one hand polarization is one of the most used degrees of freedom in quantum information protocols, widely employed both for deterministic separation of photons or to encode entanglement; on the other hand, broadband frequency states for high-dimensional encoding are attracting a growing interest thanks to their capability to convey large-scale quantum information into a single spatial mode \cite{Kues2019}, an important asset for scalability. 
Although broadband integrated polarization mode splitters working in the classical regime have been demonstrated in Si-based platforms~\cite{Wu2017,Xu2019,Guerber2018}, polymers~\cite{Huang2017} and  lithium niobate waveguide circuits~\cite{Chung2019}, to date, there exists no device able to combine the generation of broadband biphoton states and their deterministic splitting through polarization. Moreover, to combine these two very distinct functionalities in a simple monolithic design that requires only very few fabrication steps is a significant technological challenge.

Here, we demonstrate a monolithic AlGaAs waveguide circuit including the generation of broadband orthogonally polarized photon pairs via type II spontaneous parametric down conversion and their polarization splitting through a birefringent directional coupler.
AlGaAs waveguides are a promising all-rounder physical platform for quantum photonics that stands out among other systems by combining at the same time strong light confinement, high $ \chi^{(2)} $ (\SI{150}{\pico\metre\per\volt} at \SI{1550}{\nano\metre}) and electro-optic coefficient~\cite{Wang2014}, moderate propagation losses and compliance with electrical injection~\cite{Boitier2014}. A comparison between the properties of AlGaAs and other materials for quantum photonics can be found in Ref.~\cite{Appas2022}. We demonstrate that \SI{85}{\percent} of the pairs are deterministically spatially separated via their polarization over a bandwidth of \SI{60}{\nano\metre}. The performances of the device as a quantum photonic circuit are assessed by implementing a Hong-Ou-Mandel interferometer at the chip output, one of the fundamental experiments in quantum optics lying at the heart of many nonclassical logic operations; the obtained visibility is \SI{75.5}{\percent} for a \SI{60}{\nano\metre}-broad biphoton state. These results, obtained at room temperature and telecom wavelength in a material platform exhibiting valuable miniaturization and optoelectronic capabilities, represent a significant step towards real-world quantum photonic integrated circuits working in the broadband regime.

\section{Sample layout and classical characterization}

\begin{figure*}[h]
	\centering
	\includegraphics[scale=1]{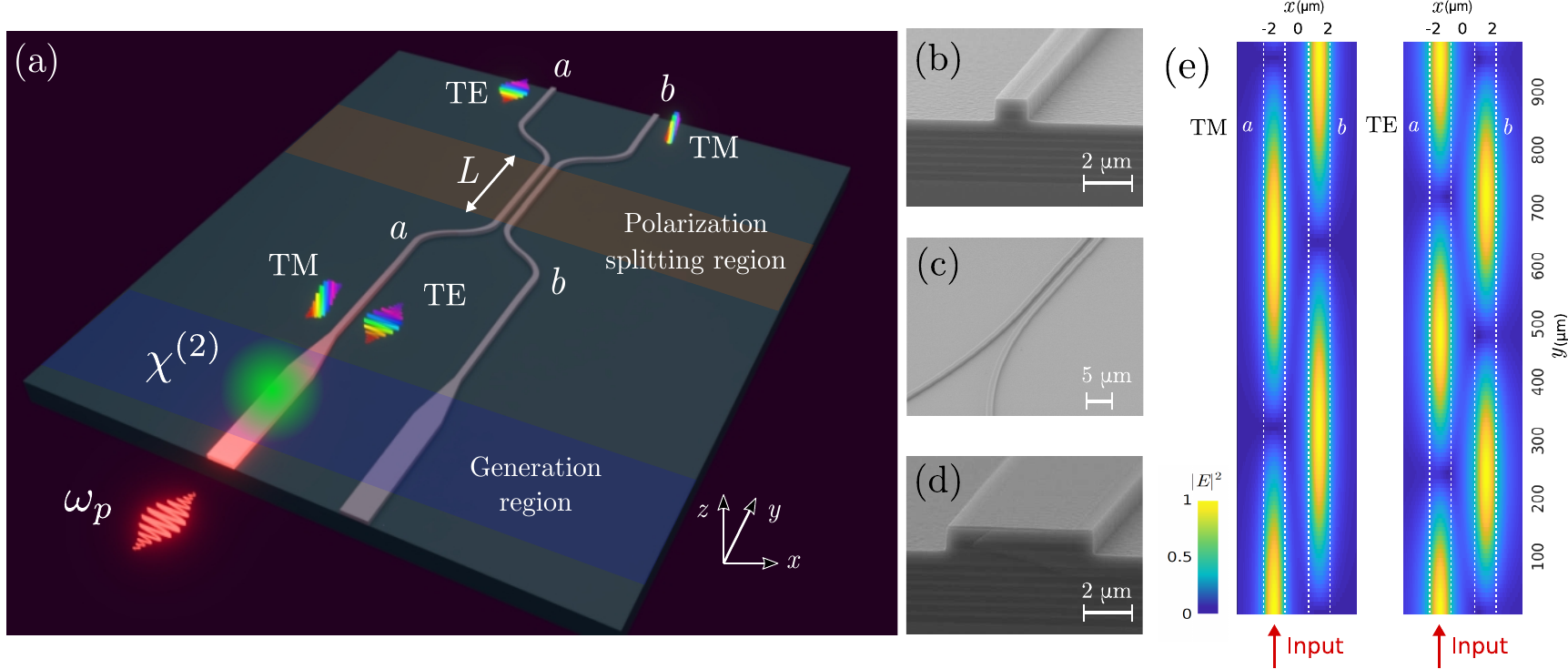}
	\caption{\label{fig:fig1} (a) Chip layout showing the photon-pair generation region and the polarization splitting region. (b-d) SEM images of the fabricated sample. (b) Output waveguide section, (c) S-Bends and polarization splitting region, (d) generation region waveguide section.
	(e) Electromagnetic simulation of the square modulus of the electric field in the polarization splitting region for light propagating in the fundamental TE and TM modes in a nominal structure.}
\end{figure*}

% general principle of the device
The device, sketched in ~\cref{fig:fig1}~(a), combines a parametric source of orthogonally-polarized photon-pairs followed by a broadband polarizing mode splitter fabricated out of an epitaxially-grown multilayer AlGaAs wafer by a one-step inductively coupled plasma etching process (see Supplementary Material for details). In this design the generation region and the polarization splitter consist of two monolithic waveguides of variable width converging at a central evanescent coupling region through S-Bend waveguides. The integration of the two functionalities on the same monolithic chip required a specific development for both the epitaxial structure and the design of the photonic circuit in order to meet the constraints related to the fabrication process and to the control on the generated quantum state as will be explained in the following.

Photon pairs in the telecom S+C band are generated into a \SI{5}{\micro\metre}-wide ridge waveguide section which is then tapered down to a narrower width $ w $ when approaching the polarization splitting region. In this part of the chip, the waveguides are separated by a gap $ g $ and evanescent coupling occurs between the guided modes of the two structures. We design the device by defining $ w $, $ g $ as well as the coupling length $ L $ such that photons generated in the TM mode, with an electric field along the $ z$-axis, are coupled into the opposite waveguide, while photons generated in the TE mode, with an electric field along the $ x$-axis, exit through the injection waveguide itself. 
%At the end of the coupling region, TE and TM photons have been split into two distinct spatially separated waveguides.
% fabrication
The device was patterned by electron-beam lithography and dry etched by inductively coupled plasma (ICP) leading to a ridge height of \SI{800}{\nano\metre} (see Supplementary Material for details). Scanning electron micrographs of the chip input, output and S-Bends regions, showing high fabrication quality with upright and smooth waveguide sidewalls, are featured in \cref{fig:fig1}~(b-d).

% working principle of the source
The generation region is based on a Bragg reflection waveguide emitting photon-pairs through Type II spontaneous parametric downconversion (SPDC)~\cite{Horn2012,Autebert2016}. We note that a specific design of the epitaxial structure of the chip has been necessary in order to cope with the fabrication issues of the splitting region. Indeed, the main drawback of the dry etching technique used to process the chip is that the etching rate is lower in the narrow gap between waveguides within the splitting region than in the rest of the chip. Numerical simulations showed that using the conventional 6-layer Bragg reflector AlGaAs waveguides described in Refs.~\cite{Autebert2016,Maltese2020} the height unbalance between the inter-waveguide gap and the outer sidewalls accumulated during the etching time would prevent reaching a high splitting ratio. We solved this issue by designing an epitaxial structure featuring the same nonlinear conversion efficiency but having only two upper Bragg reflectors ~\cite{Appas2022} (as can be seen in \cref{fig:fig1}~(b,d)) thus requiring a much shorter etch time. This results in a more uniform etching profile across the structure allowing for optimal splitting ratios.The chip is pumped by an external CW near-infrared laser. 
Typical internal pair generation rate and brightness of a \SI{2}{\milli\metre}-long straight waveguide having the same epitaxial structure are \num{7e6} pairs/s and \num{2e5} pairs/s/mW/nm (in terms of internal pump power) respectively, placing the AlGaAs platform at the state-of-the-art level for integrated sources of polarization-entangled photons~\cite{Appas2022}.
The very short length of the waveguides that stems from the high $\chi^{(2)}$ of GaAs combined with the intrinsically low group velocity mismatch between orthogonally polarized modes of the waveguides allows for the generation of photon pairs entangled in the frequency and polarization degrees of freedom without the need for off-chip temporal walk-off compensation~\cite{Autebert2016,Maltese2020,Appas2022}.
%Owing to the dispersion properties of the guided modes and the negligible group velocity mismatch between orthogonally polarized photons, this type of source allows for the generation of photon pairs entangled in the frequency and polarization degrees of freedom without the need for off-chip temporal walk-off compensation~\cite{Autebert2016,Maltese2020,Appas2021}.
Another key property of the source is its wide spectral bandwidth (\SI{60}{\nano\metre})~\cite{Appas2021} and strongly anticorrelated joint spectral amplitude~\cite{Appas2022} making it ideal for multi-user entanglement based QKD networks or high-dimensional frequency-based quantum information~\cite{Appas2021,Maltese2020}.
In this kind of structure, the SPDC phase-matching wavelength depends on the effective refractive index of the involved modes and consequently on the waveguide width. 
This property has been judiciously exploited in the design of the chip to prevent the spurious generation of photon pairs in the polarization splitting region and the output arms. By varying the waveguides width along the propagation direction throughout the device, we ensure that the pump is tuned to the phase-matching wavelength in the generation region (\SI{762.5}{\nano\metre} for the \SI{5}{\micro\metre}-wide waveguide), while it is detuned in the polarization splitting region and output arms of the device (\SI{770.5}{\nano\metre} for the \SI{1.5}{\micro\metre}-wide waveguide) and will therefore not trigger the SPDC process. 

% working principle of the polar splitting region
The polarization splitting region relies on a birefringent directional coupler architecture.
%whose design is done in the framework of the coupled-mode theory.
We denote by $ \vec{E}_a, \vec{E}_b $ the electromagnetic field of the fundamental modes of waveguides $ a $ and $b$, as indicated in~\cref{fig:fig1}~(a). Following the coupled modes theory ~\cite{AmnonYariv2006}, in the evanescent coupling region the eigenmodes of the structure take the form of symmetric $ S $ and anti-symmetric $ AS $ supermodes : $ \vec{E}_{S} = \left(\vec{E}_a + \vec{E}_b\right)/\sqrt{2} $ and $ \vec{E}_{AS} = \left(\vec{E}_a - \vec{E}_b\right)/\sqrt{2} $ with associated propagation constants $ \beta_{S} $ and $ \beta_{AS} $. As a result, light propagating along the $y$ axis in the TE (resp. TM) fundamental mode of waveguide $ a $ will hop back an forth between the two evanescently coupled waveguides with a beating length  $ \LcTE = \pi/(\beta_{S}^{\TE} - \beta_{AS}^{\TE}) $ (resp. $ \LcTM = \pi/(\beta_{S}^{\TM} - \beta_{AS}^{\TM}) $). Thanks to the intrinsic modal birefringence of the waveguides, we can find a region in the parameter space $ (w,g) $ for which:
\begin{equation}\label{eq:sweetspot}
	\dfrac{\LcTE}{\LcTM} = \dfrac{p}{p+1}.
\end{equation}
where $p$ is an integer. In our case, we have chosen $p=3$. When this condition is met, we choose a coupling length $ L = 4\LcTE $ such that incoming TE-polarized light in waveguide $ a $ is totally transferred to output port $ a $ after hopping four times between the waveguides, while TM-polarized light undergoes three hops before exiting through the opposite output port $ b $.

\begin{figure}[h]
	\centering
	\includegraphics[scale=0.65]{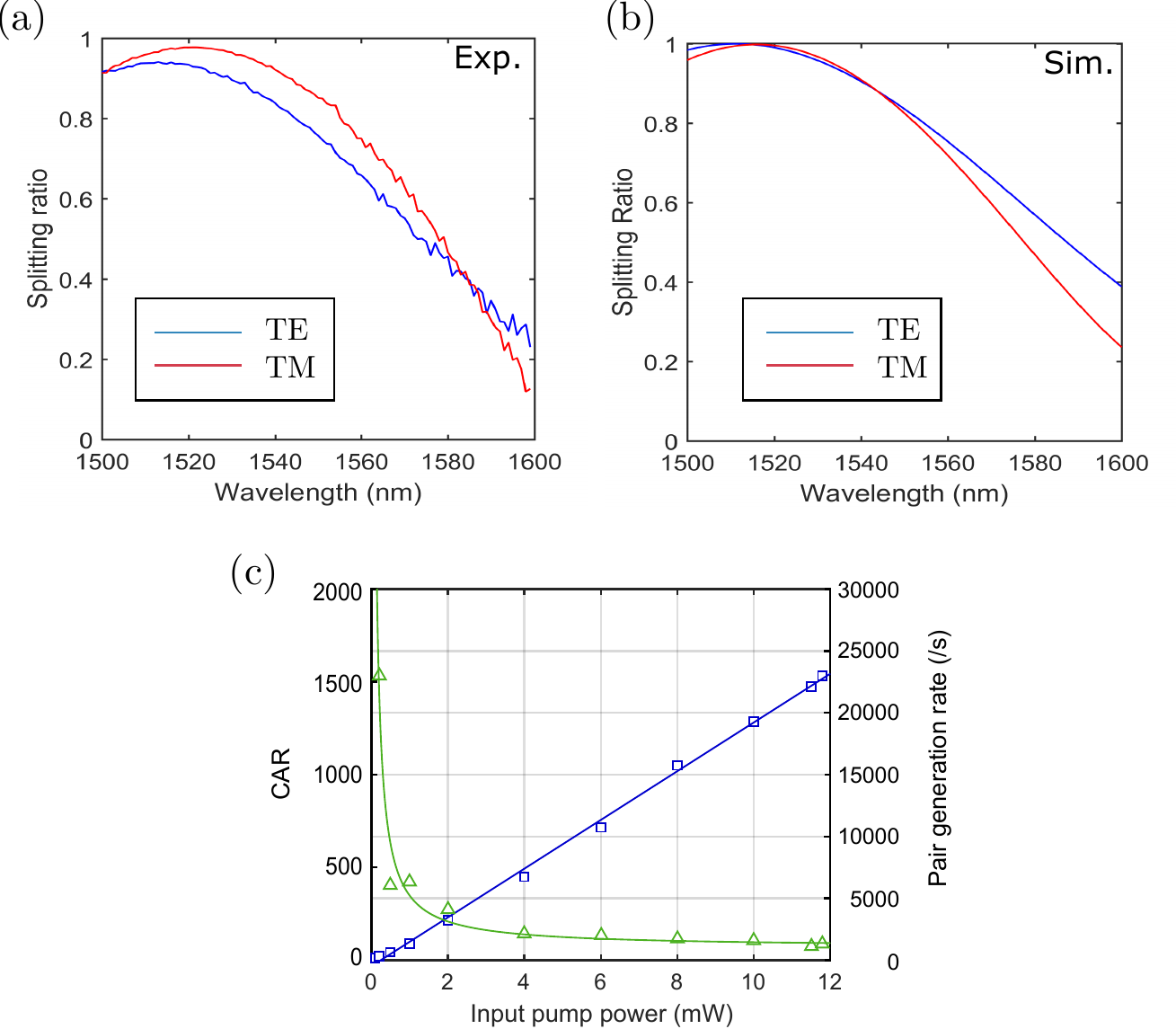}
	\caption{\label{fig:fig2} (a) Measured $ s_\TE,s_\TM $ as a function of wavelength for the chip presented in this work and (b) corresponding numerical simulations (see SM for details)	(c) Measured coincidence count rate (blue squares) and CAR (green triangles) after the output microscope objective as a function of input pump power measured before the input objective. Solid lines are a guide to the eye representing a linear trend for the count rate and an inverse law for the CAR. Errorbars derived from poissonian detection statistics are smaller than symbols.}
\end{figure}

% technical achievements in fabrication

% splitting ratio: definition, simulations
The figure of merit that we use to quantify the ability of the coupler to spatially separate the two orthogonal polarizations is the splitting ratio. For a coupling length $ L $, it is defined for each polarization as:
\begin{align}
	s_\TE &\equiv \dfrac{P^{(\TE)}_a}{P^{(\TE)}_a + P^{(\TE)}_b} = \dfrac{1}{2}\left[1+\cos\left(\pi \dfrac{L}{\LcTE}\right)\right], \label{eq:srTE} \\
	s_\TM &\equiv \dfrac{P^{(\TM)}_b}{P^{(\TM)}_a + P^{(\TM)}_b} = \dfrac{1}{2}\left[1-\cos\left(\pi \dfrac{L}{\LcTM}\right)\right], \label{eq:srTM}
\end{align}
where $ P^{\alpha}_a, P^{\alpha}_b $ is the output power at port $ a,b $ for polarization $ \alpha = \TE,\TM $.
With this definition, a perfect polarization splitting is obtained for $ s_\TE = s_\TM = 1 $ corresponding to the case where all TE-polarized light exits from arm $ a $ and all TM-polarized light exits from arm $ b $.
We perform electromagnetic simulations of the guided modes of the structure in the $ xz $ plane in the approximation of infinitely long waveguides and solve for the propagation constants $ \beta_{S}^\TE,\beta_{AS}^\TE ,\beta_{S}^\TM,\beta_{AS}^\TM $. We use these numerically calculated values to find the optimal $ (w,g) $ maximizing the splitting ratio over a spectral range matching the \SI{60}{\nano\metre} biphoton bandwidth of the source (see Supplementary Materials for details). 
%For a target central operating wavelength of \SI{1525}{\nano\metre}, corresponding to the phase-matching wavelength of the generation region, we obtain a design for which $ w_\mathrm{sim} = \SI{1.17}{\micro\metre} $, $ g_\mathrm{sim} = \SI{1.30}{\micro\metre} $ and $ L_\mathrm{sim} = \SI{988}{\micro\metre} $.
We simulate the propagation of light in the fundamental TE and TM modes by solving the Maxwell equations for the electric field using the numerically calculated propagation constants obtained for this set of parameters. The result of this simulation over a xy plane located in the waveguide core is shown in \cref{fig:fig1}~(e).
%\cref{fig:fig2}~(a) features the result of the Eigenmode Expansion (EME) simulation of the propagation of light in the fundamental TE and TM modes in the polarization splitting region for this set of parameters.
As expected, light in the TE mode exits through port $ a $ after two round trips while light in the TM mode exits through port $ b $ after one and a half round trip.
To validate our design strategy, we fabricated on a single epitaxial wafer 10 devices with fixed values of waveguide width and gap and different values of coupling length $L$. For all these devices we measured the splitting ratio for the TE and TM fundamental modes and compared them to calculations using \cref{eq:srTE,eq:srTM}.
% splitting ratio: classical charact.
%We experimentally characterize the splitting ratio by measuring the TE and TM transmission of the device and computing the splitting ratio from~\cref{eq:srTE,eq:srTM}. 

% experimental charact
The experimental characterization of the splitting ratio is done by injecting a telecom laser beam into the generation region and recording the transmitted power at the two output ports $ a $ and $ b $ of the device as a function of the input wavelength $ \lambda $. In order to access independently the TE and TM splitting ratios, the laser beam is linearly polarized either along the vertical $ z$-axis or the horizontal $ x$-axis (see \cref{fig:fig1}~(a)). Light is coupled into the waveguide using a microscope objective (NA = 0.95) and collected simultaneously from both output arms $ a,b $ of the device using a second microscope objective (NA = 0.65) before being detected with an infrared powermeter.  The chip was set at a stable temperature of \SI{20}{\degreeCelsius} using a Peltier element in a PID loop.
\cref{fig:fig2}~(a) reports the experimental results obtained for the device presented in this work, and \cref{fig:fig2}~(b) the corresponding numerical simulations, reproducing well the obtained results (see Supplementary Material (SM) for details about measurements and simulations on the complete set of 10 devices).
All subsequent experimental measurements of this paper are acquired from this device.

%We repeat this measurement on several devices having different coupling lengths $ L $. Experimental results for two representative values of $ L $ are plotted in~\cref{fig:fig2}~(c-d). We see that, in both cases, $ s_{\TE}$ and $ s_{\TM} $ reach their maximum value at approximately the same wavelength and that the bandwidth for which they are both above \SI{90}{\percent} matches the simulated value of \SI{50}{\nano\metre}.
%Moreover, we observe that by changing $ L $ the optimal wavelength can be blue or red-shifted, providing us with a useful tuning parameter to adjust the splitting ratio to the biphoton degeneracy wavelength of the source.

%The latter being measured to be \SI{1525}{\nano\metre}, we chose $ L = \SI{1080}{\micro\metre} $ (\cref{fig:fig2}~(d)) to get the best polarization splitting over the entirety of the biphoton bandwidth. 

\section{On-chip generation and polarization splitting of telecom photon pairs}

\begin{figure}[h]
	\centering
	\includegraphics[scale=0.18]{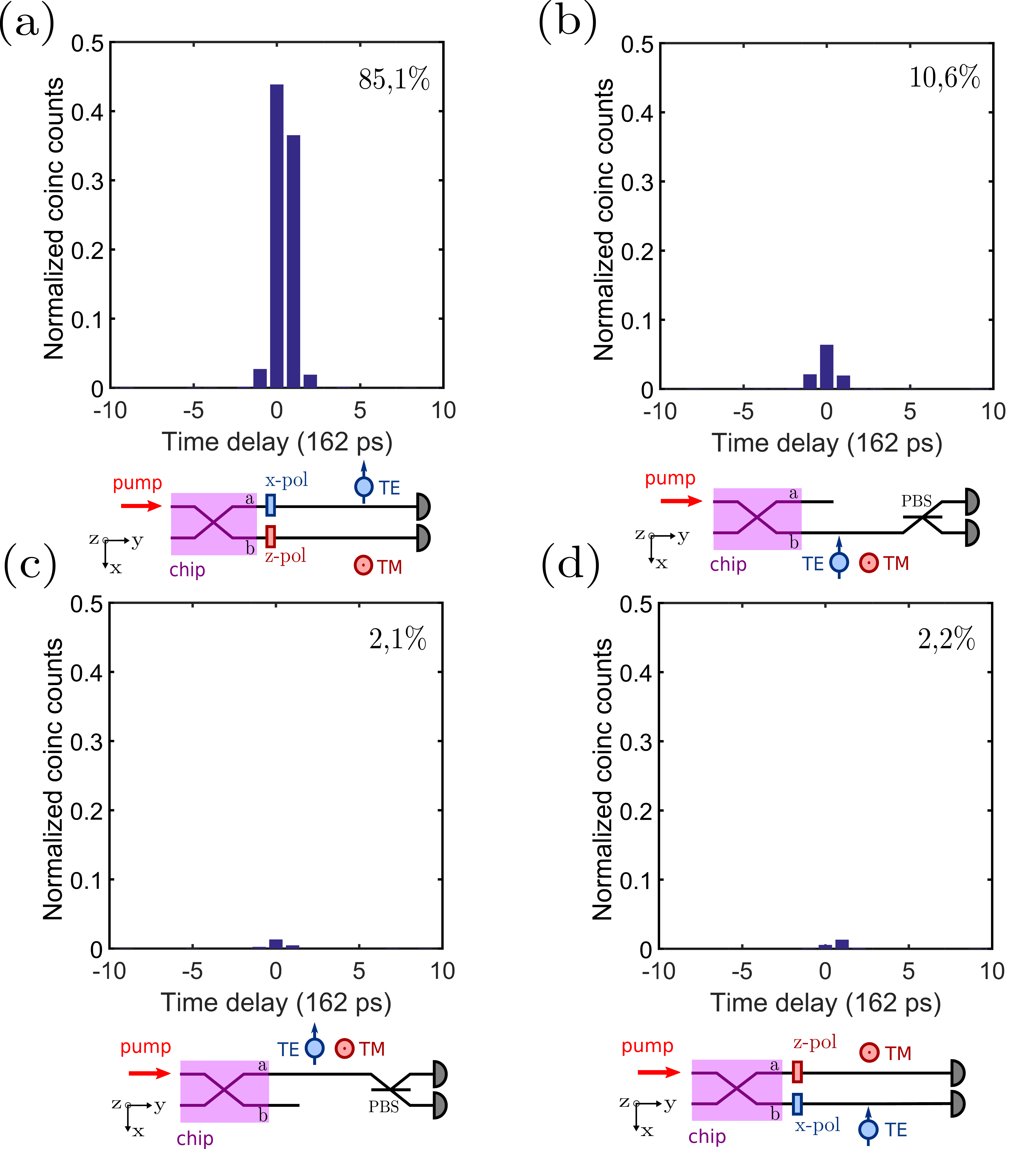}
	\caption{\label{fig:fig3} Measured coincidences normalized to the total number of coincidences measured in the following four configurations, showing the efficient on-chip generation and polarization splitting of the biphoton state : (a) Polarizer with transmission axis aligned along x in arm $ a $ and with transmission axis aligned along z in arm $ b $, (b) Arm $ a $ blocked and arm $ b $ with a PBS (c) Arm $ b $ blocked and arm $ a $ with a PBS and (d) reversed configuration with respect to (a).}
\end{figure}

We assess the performance of the device in the quantum regime by generating orthogonally polarized photon pairs via SPDC in the generation region and detecting the separated photons at the two output ports of the device. To this end, we pump the input waveguide using a near-infrared CW laser (TOPTICA DL Pro 780) of \SI{100}{\kilo\hertz} linewidth at a wavelength of \SI{762.5}{\nano\metre} using the same input and output coupling microscope objectives. Residual transmitted pump photons are filtered out using free-space long-pass filters while the generated telecom photons are fiber-coupled and detected by two superconducting nanowire single photon detectors (SNSPDs, Quantum Opus). Time-correlations between the photon detection times are recorded using a time-to-digital converter (quTools) with a \SI{162}{\pico\second} temporal resolution. The measured number of coincidence counts as a function of the pump power before the input objective and corresponding coincidence-to-accidental ratio (CAR) are displayed in~\cref{fig:fig2}~(c). We observe that
the coincidences and CAR display respectively a linear and inversely proportional dependence on the pump power as in the case of a single-waveguide AlGaAs source with off-chip pair separation~\cite{Horn2012}.
The estimated collection efficiency for each arm is \SI{11.7}{\percent} and the  SNSPD detection efficiency is \SI{85}{\percent}. The main limitation to the measured coincidence rate is then the propagation losses of \SI{0.9}{\per\centi\metre} (\SI{3.9}{\decibel\per\centi\metre}) and \SI{1.5}{\per\centi\metre} (\SI{6.5}{\decibel\per\centi\metre}) for TE and TM light respectively in the fundamental telecom mode, owing mostly to surface roughness of the waveguide and to imperfections in the materials occurring at the interface between different AlGaAs layers. The losses can be reduced in future devices by optimizing the epitaxial structure in a design involving less AlGaAs layers as well as by improving the etching recipe of the fabrication process to obtain smoother sidewalls as demonstrated in Ref.~\cite{Chang2020}.

To certify the successful polarization splitting of the generated photon pairs, we record the number of coincidences counts in four different experimental configurations (see \cref{fig:fig3}): case (a) with a linear polarizer aligned along $x(z)$ at the output of arm $a(b)$; case (b) with arm $a$ blocked and a PBS inserted before the detectors; case (c) with arm $b$ blocked and a PBS inserted before the detectors; case (d) with a linear polarizer aligned along $z(x)$ at the output of arm $a(b)$.
The configuration of case (a) selects the events corresponding to the correct functioning of the designed polarization splitter; the configurations (b)-(d) correspond to unwanted events. The results of time correlation measurements normalized to the total number of coincidences in these four situations are reported in \cref{fig:fig3}~(a)-(d). We see that the number of events measured in case a) represents \SI{85}{\percent} of the total number of emitted pairs, demonstrating the high efficiency of the polarization splitting region design. This efficiency is limited by the variation of the splitting ratio across the biphoton bandwidth, which can be reduced by the use of adiabatic couplers such as the ones described in Ref.~\cite{Chung2019}.

\section{Hong-Ou-Mandel interferometry at the chip output}

\begin{figure*}[h]
	\centering
	\includegraphics[scale=0.65]{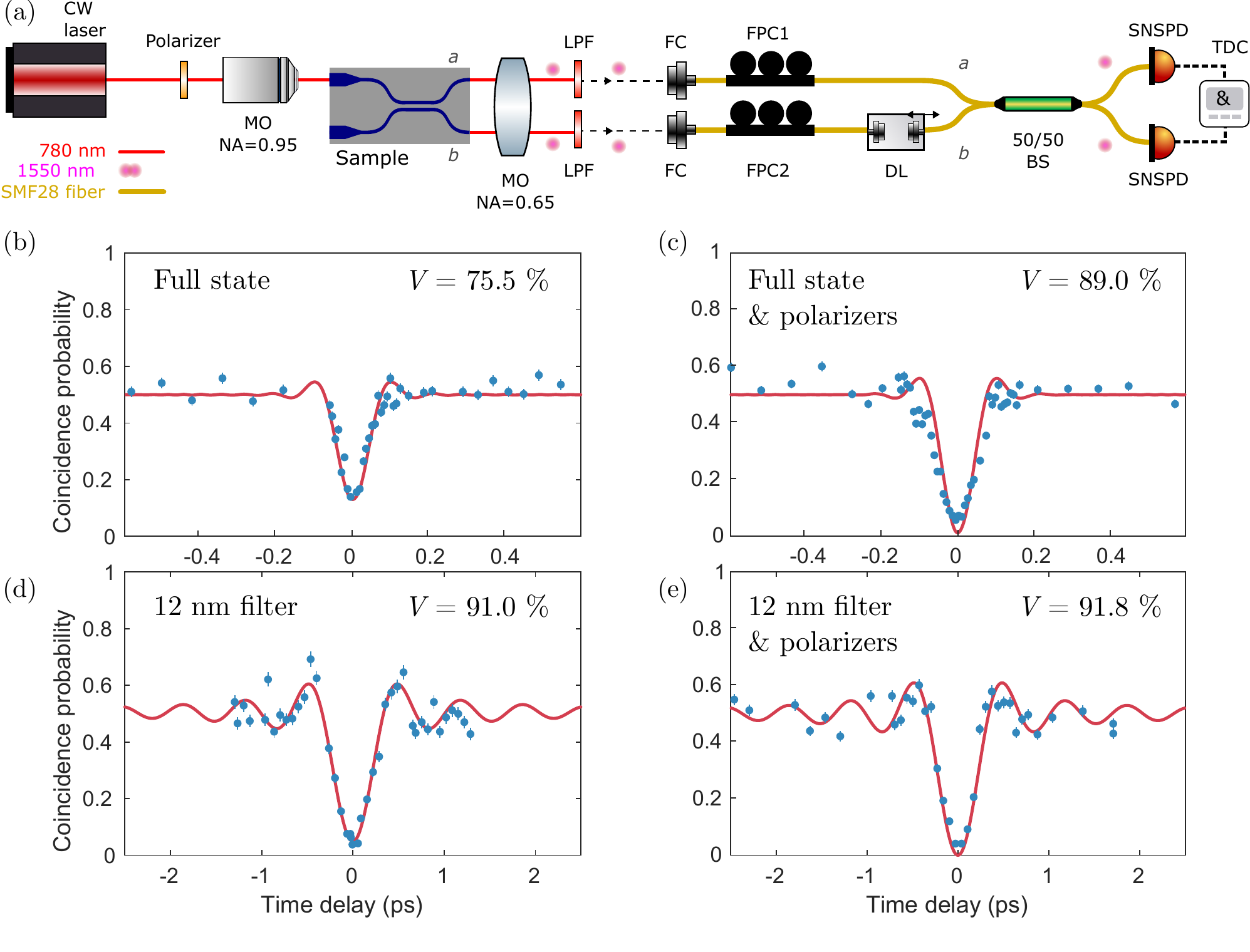}	\caption{\label{fig:fig4} (a) Experimental setup for the measurement of HOM interference at the chip output. MO: Microscope objective, LPF: Low-pass frequency filter, FC: Fiber collimator, FPC: Fiber polarization controller, DL: Delay line, SNSPD: Superconducting Nanowire Single Photon Detector, TDC: Time-to-digital converter. (b-c) Unfiltered HOM interferograms obtained (b) without and (c) with polarizers at the output of the device (see main text for details). (d-e) HOM interferograms with \SI{12}{\nano\metre} spectral filter obtained (d) without and (e) with polarizers at the output of the device. Error bars are calculated assuming Poissonian statistics.}
\end{figure*}

%\textcolor{blue}{1. Motivate HOM measurement (see review by Sciarrino and work by Nicolas and Giorgio). 2. Emphasize that there is no filtering (contrary to msmt in ppLN).}

%After showing that photon pairs can be generated and separated on-chip using the integrated polarization splitter, we probe their quantum state through Hong-Ou-Mandel (HOM) interference.
%HOM effect is intrinsically quantum and witnesses the non-classicality of generated biphoton state. In addition, this type of two-photon interference is also a cornerstone of quantum information processing widely used in quantum state preparation~\cite{Chen2018}, measurement-device independent quantum key distribution~\cite{Lo2012} or precision measurements~\cite{Hong1987}.
The device operation in the quantum regime is further probed through Hong-Ou-Mandel (HOM) interference.
The HOM effect, in addition to being a witness of non-classicality, represents a cornerstone of quantum information processing~\cite{Bouchard2021}. Indeed, this type of two-photon interference is widely used in quantum state engineering~\cite{Chen2018,Francesconi2020}, measurement-device independent quantum key distribution~\cite{Lo2012} and quantum metrology~\cite{Hong1987,Lyons2018}.
Lastly, the HOM effect has been proven to be a central tool in time-frequency encoded high-dimensional quantum information with biphoton states~\cite{Maltese2020,Fabre2020}.

%After showing that photon pairs can be generated and separated on-chip using the integrated polarization splitter, we probe their quantum state through Hong-Ou-Mandel (HOM) interference.
The experimental setup used to perform HOM interference at the output of the device is depicted in~\cref{fig:fig4}~(a). 
Input and output coupling are done with the same microscope objectives used in the previous experiment. After pump filtering and fiber collimation, we apply a temporal delay on arm $ b $ using a free-space delay line before recombining the two photons on a fiber 50/50 beam splitter (BS) and detecting them at the output of the BS with SNSPDs. The two fiber polarization controllers (FPC) are used to ensure that the polarization of the two photons is identical in the fiber interferometer.

To probe separately the impact of the properties of the generated quantum state and of the polarization splitting ratio of the device on the two-photon interference, we measure the HOM interferogram in four different settings: first, using the full generated state respectively without and with the addition of polarizers at the chip output; then in the same measurements settings but using the generated quantum state spectrally filtered around degeneracy with a 12 nm-wide filter.      

In~\cref{fig:fig4}~(b) we plot the result of a HOM measurement taken directly at the chip output without any spectral filtering nor added polarizers. The measured visibility, given by  $ V = (C_{\rm ref}-C_{\rm min})/(C_{\rm ref})$ \cite{Bouchard2021}, where $C_{\rm ref}$ is the coincidence reference level taken at large temporal delay and $C_{\rm min}$ is the coincidence minimum, is $\SI{75.5}{\percent} \pm \SI{5.4}{\percent}$ exceeding the classical threshold of \SI{50}{\percent} thus showing that the integrated source and splitter operate in the quantum regime. We stress that all the visibility values given in this work are raw, meaning that the noise counts are not subtracted. The red curve corresponds to the calculated HOM coincidence probability obtained from the simulated quantum state produced in the generation region and the experimentally measured splitting ratios $ s_\TE,s_\TM$ (see Supplementary Materials for details). We observe that the agreement between our model and the experimental results is excellent.

In~\cref{fig:fig4}~(c) we plot the results obtained by inserting in each optical paths $ a $ and $ b $ a linear polarizer with transmission axis respectively aligned along the $x$ and $z$-direction. The obtained  visibility in this case is $ V=\SI{89}{\percent} \pm \SI{4.6}{\percent} $, a value which is in agreement with previous experiments on AlGaAs sources consisting of a single straight ridge waveguide followed by an off-chip fiber polarizing beam splitter~\cite{Autebert2016,Maltese2020}, supporting the idea that this measurement gives access to the intrinsic visibility of the generated two-photon state. The limitation to this value comes from waveguide birefringence causing spectral asymmetry with respect to frequency degeneracy between TE and TM-polarized photons. As a result, our measurement shows that a perfect splitting ratio over the whole state bandwidth would allow to reach the maximum value of visibility allowed by the source properties.

We emphasize that both HOM interferograms were obtained with a broadband biphoton state as opposed to previous demonstrations reported in periodically poled lithium niobate (ppLN) where off-chip filters were used~\cite{Luo2019} or in Silicon chips generating narrowband degenerate photon pairs~\cite{He2015}.
The width of the experimentally measured HOM dip in \cref{fig:fig4}~(b-c) is approximately \SI{90}{\femto\second}, corresponding to a two-photon bandwidth of \SI{63}{\nano\metre} comparable to the expected \SI{60}{\nano\metre} reported in Ref.~\cite{Appas2021}, indicating the broadband character of the measured quantum state.
To quantify the effect of spectral filtering on the performances of our device, we repeat the HOM measurement after adding at the chip output a \SI{12}{\nano\metre} filter centered on the biphoton degeneracy frequency. We obtain a value of $ V = \SI{91.0}{\percent} \pm \SI{6.6}{\percent}$  without polarizers (\cref{fig:fig4}~(d)) and $V = \SI{91.8}{\percent} \pm \SI{6.9}{\percent}$ with polarizers (\cref{fig:fig4}~(e)). The improvement in HOM visibility when adding polarizers is negligible here since the splitting ratio stays very high in this \SI{12}{\nano\metre} bandwidth as shown in \cref{fig:fig2}~(a).  This value is of the same order as in previous state-of-the-art realizations in other platforms, but we notice that in our case the available spectral bandwidth is one order of magnitude higher (\SI{1.2}{\nano\metre} in~\cite{Luo2019}).
We stress the relevance of this result by pointing out that spectral filtering is commonly used in many applications such as wavelength-multiplexed quantum communications~\cite{Appas2021} or frequency-based quantum information~\cite{Kues2017} where the emitted photons have to be divided into spectral bins by the use of filters. Hence, in most cases, the spectral asymmetry with respect to frequency degeneracy between the two photons of the pairs causing imperfect HOM visibility is naturally suppressed by the action of filtering.

%\textcolor{blue}{1. Give some numbers about $ V $ with filtering. 2. Emphasize that in the context of quantum comms. need for filtering for wavelength multiplexing "naturally" increases the indistinguishability.}

\section{Conclusion and outlook}
In summary, we have demonstrated a monolithic nonlinear AlGaAs waveguide circuit integrating the generation of spectrally broadband photon pairs in the telecom range and their polarization splitting with a birefringent directional coupler. By precise engineering of the waveguide birefringence and phase-matching wavelengths, the challenging task of combining these two functionalities in a simple device made by a one-step etching process has been fulfilled. \SI{85}{\percent} of the pairs are deterministically separated by the chip over a \SI{60}{\nano\metre} bandwidth. Its performances in the quantum regime have been assessed via a Hong-Ou-Mandel interference experiment displaying a visibility of \SI{75.5}{\percent} (\SI{91}{\percent}) for a \SI{60}{\nano\metre} (\SI{12}{\nano\metre}) broad biphoton state.
We notice that a maximal visibility of the HOM dip requires that the splitting ratio curves for the TE and TM modes are superposed and symmetric with respect to the frequency degeneracy for the down-conversion process; a possible way to improve the performances of the present device is to adjust the waveguide width of the generation region to shift the frequency degeneracy of the generated state at the optimal wavelength. 
Further progress on the device performances is possible following two main directions: on one hand the design of the epitaxial structure could be optimized in order to reduce the modal birefringence thus increasing the HOM visibility~\cite{Chen2019}. On the other hand, the design of the polarization splitting region could be refined by using adiabatic couplers which, in LN waveguides without an on-chip photon pair source, have been shown to exhibit a flat spectral profile with splitting ratios above \SI{98}{\percent} over a spectral range of more than \SI{100}{\nano\metre}~\cite{Chung2019}. Another interesting approach could be the use of inverse design\cite{Molesky2018} to improve the figure-of-merit for the directional coupler, as done for example in Ref. \cite{10.1117/12.2210848}.
In any case, the device presented in this work can already be used as a source of broadband frequency anticorrelated photon pairs separated in two different spatial modes, directly employable for information processing protocols based on high-dimensional quantum states \cite{Imany2019,Lu2019}. Using both input arms of the splitting region to generate orthogonally polarized photon pairs would also allow to obtain a source of polarization entangled photon pairs directly separated in two spatial modes, as reported in ppLN~\cite{Sansoni2017} , albeit with a narrow bandwidth (0.25 nm) in the latter work. Using AlGaAs would not only allow a miniaturization of the device of around one order of magnitude, but also to take benefit of the low modal birefringence to avoid temporal walk-off post-compensation in frequency channel-based quantum networks. 
By taking a step further the device could be upgraded by adding an electro-optic delay line~\cite{Wang2014}, a polarization rotator ~\cite{Huan2000} and a 50/50 mode splitter~\cite{Belhassen2018}. In this configuration the photons, after being generated in a waveguide, would be deterministically separated at the polarizing mode splitter and time-shifted by the electro-optic delay line; one of them could be polarization rotated and finally the two photons would be recombined on the 50/50 mode splitter. Such nonlinear integrated quantum electro-optic circuit would enable a full flexible control over single-qubit operations~\cite{Luo2019} as well as an on-chip generation of polarization-frequency hyperentangled states~\cite{Xie2015} over a wide spectral range, opening promising perspectives for its utilization in quantum communication networks~\cite{Graham2015}.

\textbf{Note:} The authors declare no conflicts of interest.

\begin{acknowledgement}
	This work was supported by Defence Innovation Agency under grant ANR-19-ASTR- \textbf{0018 01}), the French RENATECH network,
	and by Paris Ile-de-France Region in the framework of DIM
	SIRTEQ through the Project Paris QCI and Project LION. The authors also acknowledge partial funding from the French government through the National Research Agency (ANR) in the context of the Plan France 2030 through project reference ANR-22-PETQ-0011.
	O.M. acknowledges Labex SEAM (Science and Engineering
	for Advanced Materials and devices), ANR-10-LABX-0096
	and ANR-18-IDEX-0001 for financial support.
\end{acknowledgement}

\begin{suppinfo}
	\begin{itemize}
		\item See Supplemental Material for supporting content on device fabrication, electromagnetic simulation of the splitting region and calculation of the HOM coincidence probability.
		\item Data underlying the results presented in this paper are not publicly available at this time but may be obtained from the authors upon reasonable request.
	\end{itemize}

\end{suppinfo}

\bibliography{biblio-pbs}

\end{document}

\section{Methods}
\subsection{Device epitaxial structure and fabrication}

The photonic chip is fabricated out of an epitaxially-grown AlGaAs wafer consisting of a 6-period Al$_{0.80}$Ga$_{0.20}$As/Al$_{0.25}$Ga$_{0.75}$As Bragg reflector (lower mirror), a \SI{351}{\nano\metre} Al$_{0.45}$Ga$_{0.55}$As core and a 2-period Al$_{0.25}$Ga$_{0.75}$As/Al$_{0.80}$Ga$_{0.20}$As Bragg reflector (upper mirror) grown on a GaAs wafer.
%The structure is designed such as to support both telecom fundamental guided modes and so-called near-infrared Bragg modes.
%The photonic bandgap of the two Bragg reflectors provide optical confinement of the Bragg mode.
The structure is designed such that the dispersion of the guided modes allows for Type II modal phase-matching between the so-called near-infrared TE Bragg mode at and the telecom TE and TM fundamental modes enabling the generation of cross-polarized photon pairs through Type II SPDC~\cite{Autebert2016}.
Only the first four upper layers, corresponding to the upper Bragg mirror, are etched resulting in a ridge height of around \SI{800}{\nano\metre}. The core layer is not etched to ensure optimal coupling between the propagating modes of the waveguides in the evanescent coupling region. Indeed, if the core is etched, then the overlap between the spatial profile of the fundamental modes $ \vec{E_a}, \vec{E_b} $ is drastically reduced, thus increasing the optimal coupling length and the subsequent device footprint.

The chip is fabricated using electron-beam lithography on a \SI{6}{\percent} silsesquioxane-based H-SiOx (HSQ) resist and \ce{Cl}-based ICP etching. After fabrication, the sample is cleaved mechanically along the $[1\bar{1}0]$ crystallographic axis to obtain optical quality facets. The final overall chip length, including the source and the polarization splitter, is \SI{7}{\milli\metre}. The output waveguide spacing is \SI{127}{\micro\metre}, corresponding to a standard fiber array pitch.

%Propagation losses in both wide and narrow region.

\subsection{Electromagnetic simulation of the device splitting ratio}

\begin{figure}[h]
	\centering
	\includegraphics[width=0.9\textwidth]{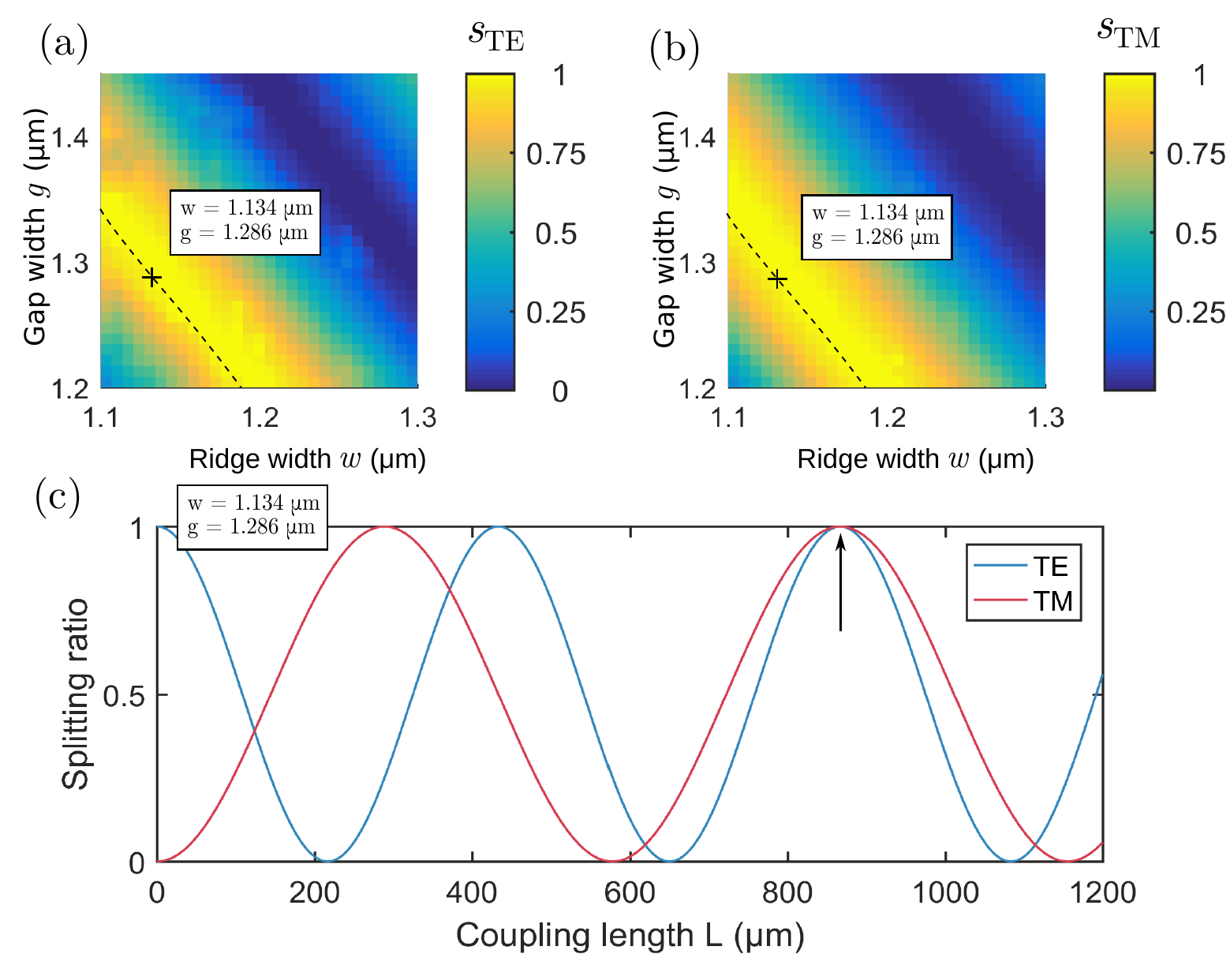}
	\caption*{Figure S1: (a) TE and (b) TM splitting ratio as a function of waveguide width $ w_\mathrm{sim} $ and gap $ g_\mathrm{sim} $ for a coupling length of $ L_\mathrm{sim} = \SI{864}{\micro\metre} $ and at a wavelength $ \lambda = \SI{1525}{\nano\metre} $. (c) Splitting ratio as a function of coupling length $ L $ for the set of parameters $ w_\mathrm{sim} = \SI{1.134}{\micro\metre}, g_\mathrm{sim} = \SI{1.286}{\micro\metre} $ and $ \lambda = \SI{1525}{\nano\metre} $. The optimal working point, for which $s_{\mathrm{TE}} = s_{\mathrm{TM}} = 1$, is indicated by an arrow.}
    \label{fig:supp1}
\end{figure}

We perform electromagnetic mode analysis in our integrated polarization mode splitter design using commercially available software suite to find a region in parameter space where condition $ \LcTE/\LcTM = 3/4 $ is fulfilled.  
%Only the first four upper layers, corresponding to two Bragg periods, are etched, leaving the core untouched. This has been done to ensure optimal coupling between the propagating modes of the waveguides. Indeed, if the core is etched, then the overlap between the spatial profile of the fundamental modes is drastically reduced, thus decreasing the coupling strength.
Maxwell's equations in the coupled waveguide structure are solved using finite element methods to obtain the spatial $ xz $ profile and effective index of a given number of modes at a fixed frequency. We implemented a simple algorithm to automatically detect the two $ S $ and $ AS $ supermodes from the multitude of guided modes that are calculated. 
%However, this procedure can feature some errors when unphysical modes are accidentally identified instead of the supermodes of interest. As a result, we obtain some outlier points in the simulation data for certain values of $ w $ and $ g $ which have been subsequently removed.

From the simulated effective $ S $ and $ AS $ mode indices, we were able to compute the transmitted power for both polarizations as a function of the waveguide width $ w_\mathrm{sim} $, gap $ g_\mathrm{sim} $ and coupling length $ L_\mathrm{sim} $ and infer the TE and TM splitting ratio as shown in Fig. S1 (a-b). 
%The regions for which $ s_\TE $ and $ s_\TM $ are 1 are highlighted with dotted lines. 
At a wavelenth of \SI{1525}{\nano\metre}, for $ L_\mathrm{sim}=\SI{864}{\micro\metre} $, we observe that the dotted lines denoting the regions where $ s_\TE $ and $ s_\TM $ are equal to 1  can be almost perfectly superimposed. As a consequence, in this design, we can achieve simultaneously perfect TE and TM polarization splitting. The optimal working point, corresponding to the region where the two TE and TM dotted lines have the best overlap, is denoted by a cross. This corresponds to a situation where $ \LcTE/\LcTM = 3/4 $.
In Fig.S1 (c), we plot the TE and TM splitting ratios as a function of coupling length for the optimal set of parameters $ w_\mathrm{sim} = \SI{1.134}{\micro\metre},\ g_\mathrm{sim} = \SI{1.286}{\micro\metre} $.
The two oscillating curves correspond to the beating of the cosine terms in~Eq.~(2-3) of the main text and the optimal coupling length $ L_\mathrm{sim}=\SI{864}{\micro\metre} $ at $\lambda = \SI{1525}{\nano\metre}$ is indicated by a black arrow.

%\subsection{Fabrication tolerance}

\section{Dependence of the splitting ratio on the coupling length $L$}

\begin{figure}[h]
    \centering
    \includegraphics[width=\textwidth]{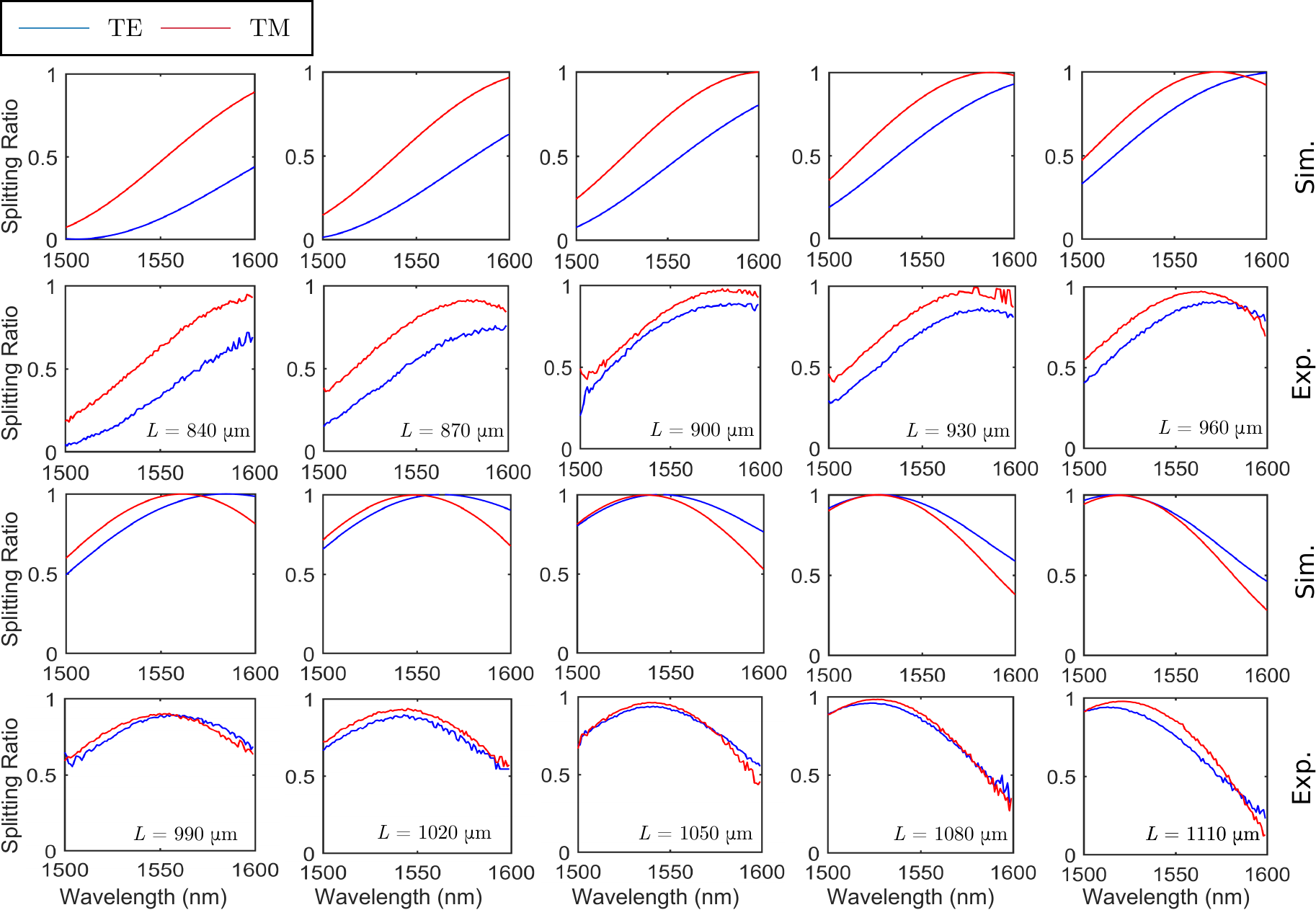}
    \caption*{Figure S2: Simulated and measured TE and TM splitting ratios as a function of wavelength for different values of the coupling length $L$ in a device with $w=\SI{1.29}{\micro\metre}$ and $g=\SI{1.51}{\micro\metre}$.}
    \label{fig:supp3}
\end{figure}

The length of the straight portion of the directional coupler provides a useful tuning parameter to adjust the splitting ratio to the biphoton degeneracy frequency. In Fig. S2 we present the measured and calculated splitting ratios for different device lengths. As can be seen the maxima of the TE and TM splitting ratios can be blue or red-shifted by decreasing or increasing $L$. The corresponding simulations were obtained with the set of parameters $ w_\mathrm{sim} = \SI{1.134}{\micro\metre},\ g_\mathrm{sim} = \SI{1.286}{\micro\metre} $ and $L_\mathrm{sim} = \gamma L $ where $\gamma = 0.8$ is a scaling factor that accounts for the effect of imperfect epitaxial structure on propagation. The experimentally measured TE and TM splitting ratios for different values of the coupling length L correspond to a device with $w=\SI{1.29}{\micro\metre}$ and $g=\SI{1.51}{\micro\metre}$.  % and $\delta=\SI{95}{\micro\metre}$ is a fixed offset that models the finite mode coupling in the input and output S-bends.

\section{Calculation of the Hong-Ou-Mandel coincidence probability}

\begin{figure}
	\centering
	\includegraphics[width=\textwidth]{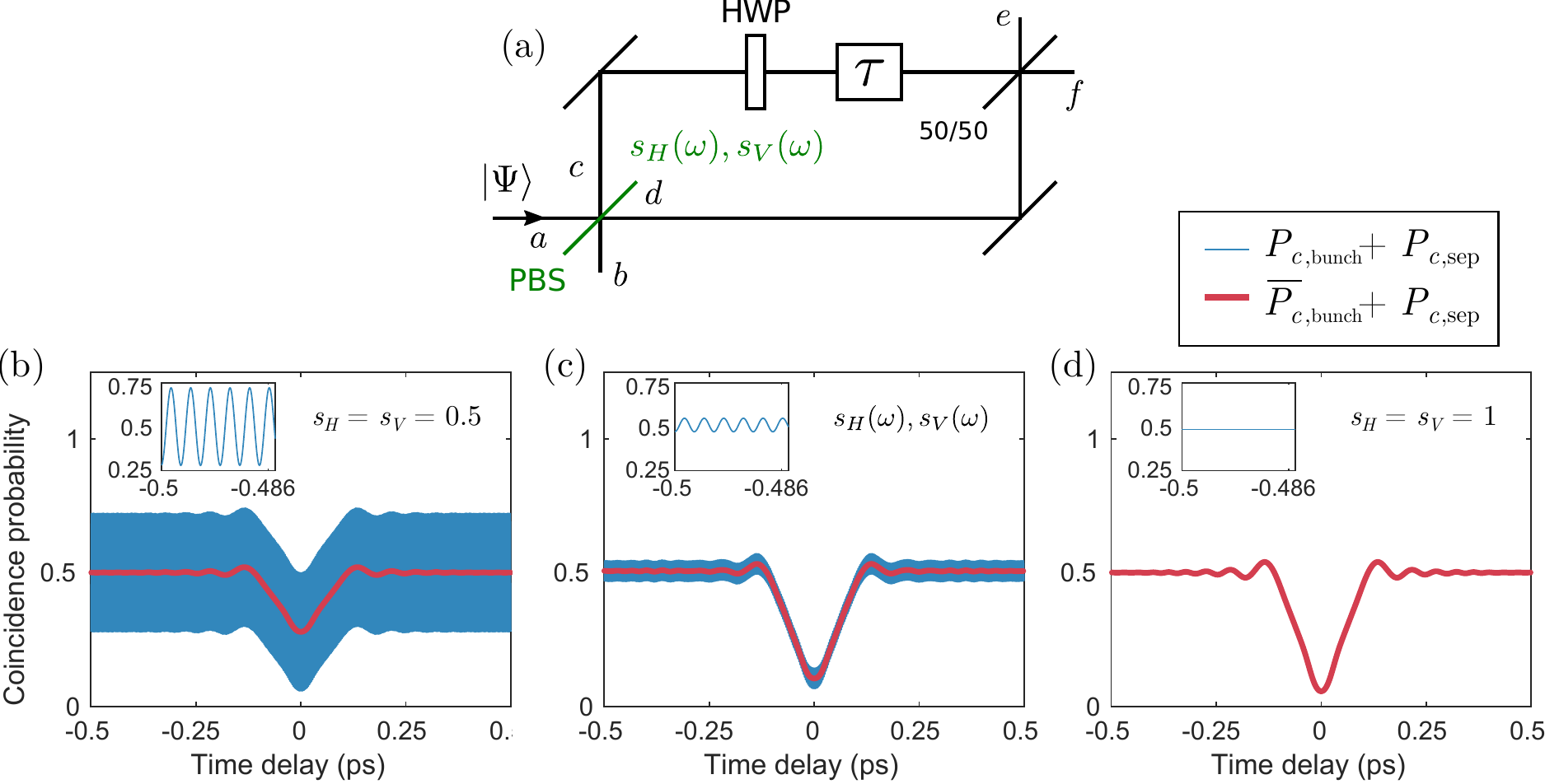}
	\caption*{Figure S3: (a) Scheme modelling a Hong-Ou-Mandel interference with a frequency-dependent polarizing beam splitter. HWP: Half-wave plate, PBS: Polarizing beam splitter. (b-d) Simulated HOM coincidence probability $ P_c $ showing different contributions from the terms $ P_{c,\mathrm{sep}} $ and $ P_{c,\mathrm{bunch}} $ for different values of the PBS splitting ratio: (b) $ s_H=s_V=0.5$, (c) $ s_H(\omega),s_V(\omega) $ given by the experimental values of the device of coupling length $ L=\SI{1080}{\micro\metre} $ (d) $ s_H=s_V=1 $. The inset displays the value of $ P_c $ over a \SI{14}{\femto\second} window at $ \tau = \SI{-0.5}{\pico\second} $ showing the oscillations originating from the term $ P_{c,\mathrm{bunch}} $.}
    \label{fig:supp2}
\end{figure}

To model Hong-Ou-Mandel (HOM) effect in our integrated device, we use the physical picture sketched in Fig. S3 (a).
The photon pairs emitted by the source are incident on a frequency-dependent polarizing beam splitter, which models the polarization splitting region of our device. We assume that the two photons are initially in the same arm $ a $ of the PBS. Then, after they exit the PBS, a delay is applied to arm $ c $ and the polarization in this arm is rotated with a half-wave plate such that the two photons end up with identical polarization. As in a usual HOM interferometer, the photons are recombined on a 50/50 beam splitter. The goal here is to calculate the coincidence probability between the output arms $ e,f $ of the 50/50 beam splitter.
We start from the generated SPDC state of our source~\cite{Appas2021}:
\begin{equation}\label{spdcstate}
	\ket{\Psi} = \iintomega \jsa \ket{\omega_1, H, a} \ket{\omega_2, V, a},
\end{equation}
where $ \ket{\omega, H,a} = \hat{a}^\dagger_H(\omega)\vacstate, \ket{\omega, V,a} = \hat{a}^\dagger_V(\omega)\vacstate $ and $ \mathcal{C}(\omega_1,\omega_2) $ the joint spectral amplitude (JSA) is assumed to be normalized $\iint d\omega_1 d\omega_2 |\jsa|^2 = 1$. Here, the labels $ H,V $ denote TE and TM modes respectively. The action of the frequency-dependent PBS is represented by a unitary transformation acting separately on the $H$ and $V$ subspaces:
\begin{align}
	\hat{a}_H(\omega) &= \sqshg\hat{c}_H(\omega) + \sqomshg\hat{d}_H(\omega), \\
	\hat{b}_H(\omega) &= \sqomshg\hat{c}_H(\omega) - \sqshg\hat{d}_H(\omega), \\
	\hat{a}_V(\omega) &= \sqomsvg\hat{c}_V(\omega) + \sqsvg\hat{d}_V(\omega), \\
	\hat{b}_V(\omega) &= \sqsvg\hat{c}_V(\omega) - \sqomsvg\hat{d}_V(\omega),
\end{align}
where the coefficients $ s_H(\omega), s_V(\omega) $ are the TE and TM splitting ratios of the polarization splitter.
After the PBS, the quantum state evolves into  $\ket{\Psi} = \ket{\Psi}_\mathrm{sep} + \ket{\Psi}_\mathrm{bunch}$ where the two independent contribution are:
\begin{align}
	\ket{\Psi}_\mathrm{sep}
	= \iintomega \jsa \ket{\omega_1, H}\ket{\omega_2, V} &\left(\sqsh \sqsv\ket{c}\ket{d} \right.\\
	&\left.+ \sqomsh \sqomsv \ket{d}\ket{c} \right), \nonumber \\
	\ket{\Psi}_\mathrm{bunch}
	= \iintomega \jsa \ket{\omega_1, H}\ket{\omega_2, V} &\left(\sqsh \sqomsv \ket{c}\ket{c} \right. \\ 
	&\left. + \sqomsh \sqsv \ket{d}\ket{d}\right). \nonumber
\end{align}
%We separated the spatial part of the kets for clarity.
These terms describe the two distinct cases where the two photons are either separated into distinct paths $\ket{\Psi}_\mathrm{sep}$ or end up into the same spatial mode $\ket{\Psi}_\mathrm{bunch}$.
After being transformed by the delay line and HWP, the state reads:
\begin{align}
	\ket{\Psi}_\mathrm{sep}
	& = \iintomega \jsa \ket{\omega_1, V, c}\ket{\omega_2, V, d} \sqsh \sqsv e^{-i\omega_1\tau}  \\
	& + \iintomega \jsa \ket{\omega_1, H, d}\ket{\omega_2, H, c}\sqomsh \sqomsv e^{-i\omega_2\tau}, \nonumber\\
	\ket{\Psi}_\mathrm{bunch}
	& = \iintomega \jsa \ket{\omega_1, V, c}\ket{\omega_2, H, c} \sqsh \sqomsv e^{-i(\omega_1+\omega_2)\tau} \nonumber \\
	& + \iintomega \jsa \ket{\omega_1, H, d}\ket{\omega_2, V, d} \sqomsh \sqsv.
\end{align}
Then the two photons enter the 50/50 beam splitter whose action can be modeled by the usual unitary transform: $\ket{c} = (\ket{e} + \ket{f})/\sqrt{2}$, $\ket{d} = (\ket{e} - \ket{f})/\sqrt{2}$. We post-select the states where both photons exit through opposite output ports and, after grouping the different terms, we obtain the following expression:
\begin{align}
	\ket{\Psi'}_\mathrm{sep}
	= \dfrac{1}{2} &\iintomega \left(\jsa \sqsh \sqsv e^{-i\omega_1\tau} \right. \nonumber \\
	& \left.- \jsat \sqsht\sqsvt e^{-i\omega_2\tau}\right) \ket{\omega_1, V, e}\ket{\omega_2, V, f} \label{eq:pbs:postsel_sep} \\
	+  \dfrac{1}{2} &\iintomega \left(\jsa \sqomsh \sqomsv e^{-i\omega_1\tau} \right. \nonumber \\
	&\left.- \jsat \sqomsht\sqomsvt e^{-i\omega_2\tau}\right) \ket{\omega_1, H, e}\ket{\omega_2, H, f}, \nonumber \\
	\ket{\Psi'}_\mathrm{bunch}
	= \dfrac{1}{2}&\iintomega \left(\jsa \sqsh \sqomsv e^{-i(\omega_1+\omega_2)\tau} \right. \nonumber \\
	&\left. - \jsat \sqomsht\sqsvt \right)\ket{\omega_1, V, e}\ket{\omega_2, H, f} \label{eq:pbs:postsel_bunch}\\
	+ \dfrac{1}{2}&\iintomega \left(\jsa \sqsh \sqomsv e^{-i(\omega_1+\omega_2)\tau} \right. \nonumber \\
	&\left. - \jsat \sqomsht\sqsvt \right)\ket{\omega_1, H, e}\ket{\omega_2, V, f}. \nonumber
\end{align}

As we can see from \cref{eq:pbs:postsel_sep,eq:pbs:postsel_bunch}, the coincidence probability will be the sum of four terms corresponding to the four possible output polarization states : $ VV,HH,VH,HV $, in other words: $ P_c = P_{c,\mathrm{sep}} + P_{c,\mathrm{bunch}} = P_{VV} + P_{HH} +  P_{VH} + P_{HV} $. The four terms can be calculated individually using the suitable projection operators~\cite{Branczyk2017b}:
\begin{equation}
	P_{\mu\nu} = \bra{\Psi} \left(\int \mtd\omega\ket{\omega,\mu,e}\bra{\omega,\mu,e}\right) 
	\left(\int \mtd\omega'\ket{\omega',\nu,f}\bra{\omega',\nu,f}\right) \ket{\Psi},
\end{equation}
with $ \mu,\nu = H,V $. We obtain the following coincidence probabilities:
%The four terms in these two equations live in orthogonal polarization subspaces $ VV$,$HH$, $VH$, $HV$. As a consequence, the associated coincidence probabilities are given by the sum of the square amplitude of those terms~\cite{Wang2006}:
\begin{align}
	P_{c,\mathrm{sep}} = P_{VV} + P_{HH}& \nonumber \\
	= \dfrac{1}{4} \iintomega |&\jsa \sqsh \sqsv e^{-i(\omega_1-\omega_2)\tau} \nonumber \\
	- &\jsat \sqsht\sqsvt|^2 \nonumber \\
	+ \dfrac{1}{4} \iintomega |&\jsa \sqomsh \sqomsv e^{-i(\omega_1-\omega_2)\tau} \nonumber \\
	- &\jsat \sqomsht\sqomsvt|^2, \\
	P_{c,\mathrm{bunch}} = P_{VH} + P_{HV}& \nonumber \\ 
	= 2 \dfrac{1}{4} \iintomega |&\jsa \sqsh \sqomsv e^{-i(\omega_1+\omega_2)\tau} \nonumber \\
	- &\jsat \sqomsht\sqsvt|^2.
\end{align}
The first term $ P_{c,\mathrm{sep}} $ defines the envelope of the HOM interferogram while the the second term $ P_{c,\mathrm{bunch}} $ is a rapidly oscillating term resulting from Franson-type interference between the two paths of the interferometer. 
%The total HOM coincidence probability is the sum of these two probabilities: $P_c = P_{c,\mathrm{bunch}} + P_{c,\mathrm{sep}}$. 
In Fig. S3 (b-d), we display the simulated coincidence probability for different values of the TE and TM ($ H $ and $ V $) splitting ratios.
%The two special cases for which $ s_H=s_V=0.5 $ and $ s_H=s_V=1 $ are displayed in~\cref{fig:supp2}~(b) and (d) respectively. The simulated HOM interferogram that is obtained by using the experimentally measured value of $ s_H(\omega), s_V(\omega) $ of our device is shown in~\cref{fig:supp2}~(c).
%The solid blue line the simulated HOM interferogram that is obtained by replacing  $ s_H(\omega), s_V(\omega) $ with the experimentally measured splitting ratios. 
%We observe that in all three cases the general shape of the interferogram follows a typical HOM dip. 
%However, we notice that there is an oscillating background, as shown in the inset, coming from the rapid oscillations of $ P_{c,\mathrm{bunch}} $. 
We observe that in all three cases the general shape of the interferogram follows a typical HOM dip with a background oscillating at a temporal period of \SI{3}{\femto\second}, which cannot be resolved by our free-space delay line. Hence when measuring the experimental HOM coincidence probability $ P_{c,\mathrm{exp}} $, these oscillations reduce to a constant background given by the average value of the oscillating term over a temporal window of \SI{30}{\femto\second} defined by the resolution of our free-space delay line :
$ P_{c,\mathrm{exp}} = \bar{P}_{c,\mathrm{bunch}} + P_{c,\mathrm{sep}} $. 
The simulated value for $ P_{c,\mathrm{exp}} $ is shown in Fig. S3 (b-d) as a solid red line. 
We note that, in agreement with experimental results, the visibility drops from $ V=\SI{89}{\percent} $ to $ V=\SI{80}{\percent} $ when going from $ s_H=s_V=1 $ (Fig. S3 (d)) to the experimental values $ s_H(\omega),s_V(\omega) $ of the splitting ratios (Fig. S3 (c)). As can be seen from our model this decrease can be attributed to the imperfect polarization splitting which gives a non-zero contribution to $ P_{c,\mathrm{bunch}} $ subsequently lowering the HOM visibility.

%We see that if $ s_H=s_V=0.5 $ then the PBS reduces to a simple 50/50 beam splitter and the HOM dip visibility cannot exceed the classical threshold of $ \SI{50}{\percent} $. 
%In the case of a perfectly polarizing PBS for which $ s_H=s_V=1 $ we have $ P_{c,\mathrm{bunch}}=0 $ and we obtain the maximum possible visibility, here $ \SI{89}{\percent} $ which is then only limited by the birefringence of the source. For the intermediate case of our device with $ L=\SI{1080}{\micro\metre} $, where $ s_H(\omega),s_V(\omega) $ follow the experimentally measured profile displayed in Fig.~2~(d) of the main text, we see that the HOM visibility is slightly lower and equal to \SI{80}{\percent}. This decrease can be attributed to the imperfect polarization splitting which gives a non-zero contribution to $ P_{c,\mathrm{bunch}} $.

\bibliography{biblio-pbs}